\documentclass[aps,prb,twocolumn,showpacs,preprintnumbers,amsmath,amssymb,superscriptaddress]{revtex4-1}

\usepackage{graphicx}% Include figure files
\usepackage{dcolumn}% Align table columns on decimal point
\usepackage{bm}% bold math

\begin{document}

\title{Beating in electronic transport through quantum dot based devices}

\author{Piotr Trocha}
\email{ptrocha@amu.edu.pl}\affiliation{Department of Physics,
Adam Mickiewicz University, 61-614 Pozna\'n, Poland}

\date{\today}

\begin{abstract}
Electronic transport through a two-level system driven by external
electric field and coupled to (magnetic or non-magnetic) electron
reservoirs is considered theoretically. The basic transport
characteristics such as charge and spin current and tunnel magnetoresistance (TMR)
are calculated in the weak coupling approximation by the use of
rate equation connected with Green function formalism and
slave-boson approach. The time dependent phenomenon is considered
in the gradient expansion approximation. The results show that
coherent beating pattern can be observed both in current and TMR.
The proposed system consisting of two quantum dots attached to
external leads, in which the dots' levels can be tuned
independently, can be realized experimentally to test this well
known physical phenomenon. Finally, we also indicate possible practical
applications of such device.
\end{abstract}


\maketitle

\section{Introduction}\label{Sec:1}
Beating is a well known phenomenon in physics.\cite{fishbane} It
occurs when the difference between frequencies of two interfering
waves is small enough. As a result a long-wavelength pattern appears
(with characteristic envelope changing very slowly). The resulting
beating frequency is equal to one half of the difference in original
wave frequencies. This effect is very important from both the
fundamental and application points of view, as it provides a
sensitive method for measuring the frequency difference. In music,
for instance,  the beating effect is used for tuning the
instruments. This phenomenon is also utilized in conventional
electronics to change the frequency of the input signal (in so
called down conversion), which helps to improve sensitivity and
selectivity of a receiver. Beating effect is also used in microwave
spectroscopy.\cite{qin}

Recently, it has turned out that beating phenomenon can be
observed in different quantum systems like for instance, a single quantum
dot (QD).\cite{gupta} Discreteness of dot's energy levels arising from quantum confinement
make it able to mimic behavior of real atom, and is thus frequently
referred to as artificial atom.\cite{reimann}
Moreover, beating has also been reported in a qubit coupled to a fluctuator being in
contact with a heat bath.\cite{galperin,loss} The beating in Rabi
oscillations\cite{simmonds,koppens} were noticed, when the
fluctuator is close to resonance with the qubit and the damping is
weak enough.\cite{galperin} Coherent beating in the
magnetoresistance of ballistic tunnel junctions were also
investigated.\cite{euges}

The beating phenomenon in the occupation probability of excited
state of a qubit has been predicted for Josephson qubit coupled
resonantly to a two-level system (TLS), (i.e., the qubit and TLS
have equal energy splittings).\cite{ku} However, this was only
true when there was any source of decoherence. This is also why
the beating phenomenon has not yet been experimentally verified in
such a system. In turn, control of electron spin coherence in
quantum dots may be provided, for instance by circularly polarized
laser pulses. Consequently, quantum dots may enable us to observe
beating. In fact, the beating have already been noticed in a few
experiments exploring time dependent Faraday
rotation\cite{wieman,gupta,greilich} in self-assembled QDs systems.

However, the beating phenomenon in electronic transport through
laterally confined quantum dots systems is an unexplored field. Moreover, there is no
experiment showing beating in transport characteristics of such
nanoscale devices. So far, investigations were mainly
focused on the spin-independent case, where only the coherent
oscillations were reported.\cite{jauho93,jauho94} Recently, Souza
has shown that the coherent oscillations become spin dependent
when Zeeman splitting of the dot's level and/or ferromagnetic
leads are considered.\cite{souza} In this case, the two spin
components of the current oscillate with different frequencies and
the beating is reported for relatively small splitting in the
frequencies (i.e., dot's level). Moreover, Prefetto \emph{et al.} have shown that intradot
spin-flip scattering suppresses the amplitude of the beating.\cite{stefanucci2}
Recently, beating in current have been predicted due to the presence of Andreev bound states
in dc biased QD system (coupled to superconducting leads) and
irradiated with a microwave field of appropriate frequency.\cite{stefanucci3}
More recently, the beating phenomenon in
coherent transport through  a microscale back-gated substrate
coupled to optically gated quantum dot has been predicted when
the Rabi frequencies approach the intrinsic Bohr frequencies in
the dot.\cite{walczak}

Here, we propose another quantum system, where the beating
can be observed. Especially, we consider two single-level quantum
dots attached to ferromagnetic/nonmagnetic leads or to spin batteries.
Experimentally it can be fabricated making use of a two-dimensional electron
gas formed at the interface of semiconductor heterostructure.
The system is
designed in specific way to avoid the channel mixing effects
between the dots. Thus, the indirect coupling between the dots is
eliminated. Moreover, the direct hopping is also excluded and the
dots can be treated as independent. Charge, spin
current and tunnel magnetoresistance (TMR) are
derived in the weak coupling approximation utilizing rate equation
associated with the Green function formalism as well as within the
slave-boson approach.\cite{dong,dong2} The gradient expansion is
utilized to include time-dependent phenomenon.

The paper is organized as follows. In section \ref{Sec:2} we
describe the model and theoretical formalism. Numerical results on
current and TMR are presented and discussed in section
\ref{Sec:3}. Summary and final conclusions are gathered in section
\ref{Sec:4}.

\section{Model and theoretical formalism}\label{Sec:2}
\begin{figure}[t]
\begin{center}
  \includegraphics[width=0.6\columnwidth]{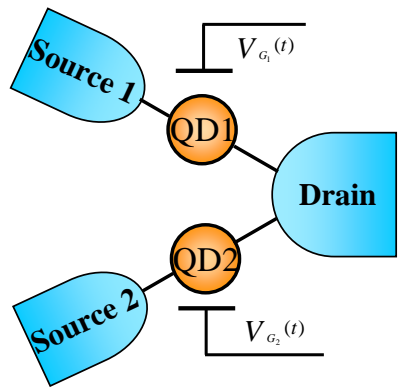}
  \caption{\label{Fig:1}
Schematic picture of two dots coupled to external leads. Each dot
is attached to its own source lead, whereas the drain electrode is
the same for the two dots.
}
\end{center}
\end{figure}

We consider two single-level quantum dots coupled to external
electrodes (magnetic or nonmagnetic). Moreover, nonmagnetic leads can be driven by both
charge and spin bias voltage.
The quantum dots are
attached to the leads as shown in Fig.\ref{Fig:1}. As channel mixing effects\cite{trocha} are
minimized, we are allowed to
introduce two independent transport channels, provided some
additional assumptions are also made. Specifically, we also
eliminate direct hopping between the dots (by creation of
sufficiently wide and high tunnel barrier between them).
The indirect coupling may
be significantly reduced in comparison to dot-lead coupling
when, for instance destructive interference effects take place.
In real systems such processes are present leading to suppression of the channel mixing effects.
As the interdot Coulomb interactions are at least an order of magnitude
smaller than the intradot Coulomb interactions, we omit the
former.

Then Hamiltonian of the system is as follows:
\begin{align}\label{eq1}
\hat{H}&=\sum_{\mathbf k\alpha\sigma} \varepsilon_{{\mathbf
k}\alpha \sigma}c^{\dagger}_{{\mathbf k}\alpha \sigma} c_{{\mathbf
k}\alpha \sigma}+
\sum_{i=1,2}\sum_{\sigma}\limits\epsilon_{i\sigma}(t)q^\dag_{i\sigma}q_{i\sigma}
       \nonumber\\
&+\sum_{i=1,2}\limits U_in_{i\sigma}n_{i\bar{\sigma}}+\sum_{{\mathbf
k}\alpha}\sum_{i,\sigma}\limits(V_{i\sigma}^\alpha
   c^\dag_{{\mathbf k}\alpha\sigma}q_{i\sigma}+\rm H.c.).\ \
\end{align}
The first term describes here the three leads in the
non-interacting quasi-particle approximation, where $\alpha
=S1,S2,D$ means two sources and one drain leads. Here,
$c^{\dagger}_{\mathbf{k}\alpha\sigma}$
($c_{\mathbf{k}\alpha\sigma}$) is the creation (annihilation)
operator of an electron with the wave vector $\mathbf{k}$ and spin
$\sigma$ in the lead $\alpha$, whereas
$\varepsilon_{\mathbf{k}\alpha\sigma}$ denotes the corresponding
single-particle energy. The next two terms in the Hamiltonian
(\ref{eq1}) describe the two quantum dots. Here,
$n_{i\sigma}=q^\dag_{i\sigma}q_{i\sigma}$ is the particle number
operator ($i=1,2$, $\sigma=\uparrow ,\downarrow$),
$\epsilon_{i\sigma}(t)$ is the discrete energy level of the $i$-th
dot (including time dependence of the corresponding gate voltage),
 and $U_i$ is the intra-dot Coulomb integral.
The last term of Hamiltonian (\ref{eq1}) describes
electron tunneling between the leads and dots, where
$V_{i\sigma}^\alpha$ are the relevant tunneling matrix elements.
Coupling of the dots to external leads can be parameterized in
terms of $\Gamma^\alpha_{i\sigma}(\epsilon)=2\pi\sum_\mathbf{k}
V_{i\sigma}^\alpha
V^{\alpha\ast}_{i\sigma}\delta(\epsilon-\varepsilon_{\mathbf{k}\alpha
i})$. We assume that $\Gamma^\alpha_{i\sigma}$ is constant within
the electron band,
$\Gamma^\alpha_{i\sigma}(\epsilon)=\Gamma^\alpha_{i\sigma}={\rm
const}$ for $\epsilon\in\langle-W/2,W/2\rangle$, and
$\Gamma^\alpha_{i\sigma}(\epsilon)=0$ otherwise. Here, $W$ denotes
the electron band width.

As in our model the dots are independent -- they do not interact
with each other -- one can decompose the density matrix operator
of the whole system as follows
$\hat{\rho}_{total}=\hat{\rho}_1\otimes\hat{\rho}_2$ and consider
each subsystem (each dot coupled to the source and drain leads)
separately.

Furthermore, we adopt the formalism presented in Ref.[\onlinecite{dong}].
Specifically, we express i-th dot's operator in terms of Hubbard
operators\cite{hubbard}
%\begin{eqnarray}\label{eq2}
%q_{i\sigma}=|0\rangle_{ii}\langle\sigma|+\sigma|\bar{\sigma}\rangle_{ii}\langle
%2|
%\end{eqnarray}
represented by four possible electron states in each
dot\cite{anderson} which satisfied the corresponding completeness relations.\cite{dong}
In the next step, the set of auxiliary operators is introduced and the dots' operators are
expressed by means of these slave-boson and pseudofermion operators.
%
%
%In further considerations we omit the dot's
%index $i$ as the further equations for both QD's are the same.
%The completeness relation is:
%
%\begin{equation}\label{eq3}
%|0\rangle\langle
%0|+\sum_{\sigma}\limits|\sigma\rangle\langle\sigma|+|2\rangle\langle
%2|=\hat{1}.
%\end{equation}
%
%When introducing the new set of operators defined in the following
%way: $|0\rangle = e^{\dag}, |\sigma\rangle = f_\sigma^\dag ,
%|2\rangle =d^\dag$,  Eq.(\ref{eq2}) becomes:
%
%\begin{align}\label{eq5}
%q_\sigma & =e^{\dag}f_\sigma+\sigma f_{\bar{\sigma}}^{\dag}d,
%\\
% e^{\dag}e&+\sum_{\sigma}\limits
% f_{\sigma}^{\dag}f_{\sigma}+d^{\dag}d=\hat{1}
%\end{align}
%
%%%%%%%%%%%%%%%%%%%%%%%%%%%%%%%%%%%%%%%%%%%%%%%%
%Slave-boson operators
%Here, $b^{\dag}$ is the slave-boson operator which creates an
%empty state, $f^{\dag}_{\sigma}$ is a peudo-fermion operator which
%creates a singly occupied state with an electron with spin
%$\sigma$, whereas $d^{\dag}$ creates doubly occupied state with an
%electron with spin $\sigma$ and other electron with spin
%$\bar{\sigma}$.
%
%In this auxiliary particles representation the
%dot's occupation operator is expressed as follows
%$n_{\sigma}=f_{\sigma}^{\dag}f_{\sigma}+d^{\dag}d$.
From the
definitions of the Dirac brackets one is able to find the commutations (and
anticommutations) rules for new operators.\cite{guillou}
%
%\begin{align}\label{eq6}
% ee^{\dag}&=dd^{\dag}=1,\ \ \ \ \ \ \ \  f_{\sigma}f_{\sigma '}^{\dag}=\delta_{\sigma\sigma'}
% \\
% ed^{\dag}&=ef_{\sigma}^\dag=f_{\sigma}e^\dag=f_{\sigma}d^\dag=de^\dag=df_{\sigma}^{\dag}=0
%\end{align}
%
Therefore, the Hamiltonian of the system  acquires the form:
\begin{align}\label{eq7}
\hat{H}&=\sum_{\mathbf k\alpha\sigma} \varepsilon_{{\mathbf
k}\alpha \sigma}c^{\dagger}_{{\mathbf k}\alpha \sigma} c_{{\mathbf
k}\alpha \sigma}+
\sum_{i\sigma}\limits\epsilon_{i\sigma}(t)(f^\dag_{i\sigma}f_{i\sigma}
   + d_i^{\dag}d_i)
      \nonumber \\
   &+ \sum_i\limits U_id_i^{\dag}d_i
   + \sum_{{\mathbf k}\alpha}\sum_{i\sigma}\limits [V_{i\sigma}^\alpha
   c^\dag_{{\mathbf k}\alpha\sigma}(e_i^{\dag}f_{i\sigma}+\sigma f_{i\bar{\sigma}}^{\dag}d_i)
   +\rm H.c.].
   \end{align}
Here, $b_{i}^{\dag}$ is the slave-boson operator which creates an
empty state in $i$th dot, $f^{\dag}_{i\sigma}$ is a peudo-fermion operator which
creates a singly occupied state with an electron with spin
$\sigma$, whereas $d_{i}^{\dag}$ creates doubly occupied state with an
electron with spin $\sigma$ and other electron with spin
$\bar{\sigma}$ in $i$th dot.

In the slave-particle representation the density matrix elements
(for each subsystem) are written in the following way:
$\hat{\rho}_{00}^{i}=e_{i}^{\dag}e_{i}$,
$\hat{\rho}_{\sigma\sigma}^{i}=f_{i\sigma}^{\dag}f_{i\sigma}$,
$\hat{\rho}_{22}^{i}=d_{i}^{\dag}d_{i}$. Here, the statistical expectations of
the density matrix elements ($\rho_{nn}^{i}\equiv\langle\hat{\rho}_{nn}^{i}\rangle$ with $n=0, \sigma, 2$) give the occupation
probabilities of the given quantum dot being empty, singly
occupied by electron with spin-$\sigma$, or doubly occupied,
respectively.

To derive the rate equations we start from the von Neumann
equation for density matrix operator:
\begin{eqnarray}\label{eq9}
 \dot{\hat{\rho}}_j=i[H,\hat{\rho}_j],
\end{eqnarray}
where
$\hat{\rho}_j=(\hat{\rho}_{00}^{j},\hat{\rho}_{\uparrow\uparrow}^{j},\hat{\rho}_{\downarrow\downarrow}^{j},
\hat{\rho}_{22}^{j})^T$ with $j=1,2$.
%
%Introducing the dot-lead Green functions
%
%\begin{align}\label{eq10}
%G^<_{e\sigma,k\alpha\sigma'}(t,t')&=i\langle c^{\dagger}_{{\mathbf
%k}\alpha \sigma'}(t')e^{\dag}(t)f_\sigma (t)\rangle,
%\notag \\
%G^<_{d\sigma,k\alpha\sigma'}(t,t')&=i\langle c^{\dagger}_{{\mathbf
%k}\alpha \sigma'}(t')f_\sigma^{\dag}(t)d(t)\rangle,
%\nonumber \\
%G^<_{k\alpha\sigma',e\sigma}(t,t')&=i\langle
%f_\sigma^{\dag}(t')e(t')c_{{\mathbf k}\alpha \sigma'}(t)\rangle,
%\nonumber \\
%G^{<}_{k\alpha\sigma',d\sigma}(t,t')&=i\langle
%d^{\dag}(t')f_{\sigma}(t')c_{{\mathbf k}\alpha \sigma'}(t)\rangle,
%\end{align}
%
The obtained averaged equations for density matrix elements can be
expressed by means of dot-lead Green functions.
%\begin{align}\label{eq11}
% \dot{\rho}_{00}=&-\sum_{k\alpha\sigma}[V_{\alpha\sigma}G^<_{e\sigma,k\alpha\sigma}(t,t)-
% V_{\alpha\sigma}^{\ast}G^<_{k\alpha\sigma,e\sigma}(t,t)],
% \nonumber \\
% \dot{\rho}_{\sigma\sigma}=&
% \sum_{k\alpha}[V_{\alpha\sigma}G^<_{e\sigma,k\alpha\sigma}(t,t)-V_{\alpha\sigma}^{\ast}
% G^<_{k\alpha\sigma ,e\sigma}(t,t)-
%\nonumber \\
% &\bar{\sigma}V_{\alpha\bar{\sigma}}
% G^<_{d\sigma,k\alpha\bar{\sigma}}(t,t)+
% \bar{\sigma}V_{\alpha\bar{\sigma}}^{\ast}
% G^<_{k\alpha\bar{\sigma},d\sigma}(t,t)],
% \nonumber \\
% \dot{\rho}_{22}=&
% \sum_{k\alpha\sigma}\sigma[V_{\alpha\sigma}G^<_{d\bar{\sigma},k\alpha\sigma}(t,t)-
% V_{\alpha\sigma}^{\ast}G^<_{k\alpha\sigma,d\bar{\sigma}}(t,t)].
% \end{align}
%
Furthermore, using Langreth theorem\cite{jauho93}, we express the dot-lead
Green functions by means of dot's Green functions and free leads'
Green functions. After utilizing gradient expansion approximation
these Green functions can be written in the $\omega$ space in the
following way:
\begin{align}\label{eq12}
G^<_{k\alpha\sigma,e\sigma}(\omega,\bar{t})&=V_{\alpha\sigma}[g_{k\alpha\sigma}^{r}G_{e\sigma\sigma}^{<}(\omega,\bar{t})+
g_{k\alpha\sigma}^{<}G_{e\sigma\sigma}^{a}(\omega,\bar{t})],
\nonumber \\
G^<_{e\sigma,k\alpha\sigma}(\omega,\bar{t})&=V_{\alpha\sigma}^{\ast}
[G_{e\sigma\sigma}^{r}(\omega,\bar{t})g_{k\alpha\sigma}^{<}+
G_{e\sigma\sigma}^{<}(\omega,\bar{t})g_{k\alpha\sigma}^{a}],
\nonumber \\
G^{<}_{k\alpha\sigma,d{\bar\sigma}}(\omega,\bar{t})&=V_{\alpha\sigma}[g_{k\alpha\sigma}^{r}
G_{d\bar{\sigma}\bar{\sigma}}^{<}(\omega,\bar{t})+
g_{k\alpha\sigma}^{<}G_{d\bar{\sigma}\bar{\sigma}}^{a}(\omega,\bar{t})],
\nonumber \\
G^<_{d\sigma,k\alpha\sigma}(\omega,\bar{t})&=V_{\alpha\sigma}^{\ast}
[G_{d\bar{\sigma}\bar{\sigma}}^{r}(\omega,\bar{t})g_{k\alpha\sigma}^{<}+
G_{d\bar{\sigma}\bar{\sigma}}^{<}(\omega,\bar{t})g_{k\alpha\sigma}^{a}],
\end{align}
where the free leads lesser ($<$), retarded ($r$) and advanced ($a$)
Green functions have the following form:
\begin{align}\label{eq13}
g_{k\alpha\sigma}^{<}=&i2\pi
f^{\alpha}(\omega)\delta(\omega-\varepsilon_{{\mathbf k}\alpha
\sigma})
\\
g_{k\alpha\sigma}^{r,a}=&P\left(\frac{1}{{\omega}-\varepsilon_{{\mathbf
k}\alpha\sigma}}\right)\mp i\pi\delta(\omega-\varepsilon_{{\mathbf
k}\alpha \sigma}),
\end{align}
with $f^{\alpha}(\omega)$ being Fermi-Dirac function for the $\alpha$ lead.
In the above equation and in further considerations we omit the dot's
index $i$ as the further equations for both QD's acquire the same form.
In Eq.(\ref{eq12}) the Green functions of the dot, in time space, are defined
as: $G_{\sigma\sigma}(t,t')=\langle\langle
q_{\sigma}(t)|q_{\sigma}^{\dag}(t')\rangle\rangle=\langle\langle
e_{\sigma}^{\dag}(t)f_{\sigma}(t)|f_{\sigma}^{\dag}(t')e(t')\rangle\rangle+|\sigma|^2\langle\langle
f_{\bar{\sigma}}^{\dag}(t)d(t)|d^{\dag}(t')f_{\bar{\sigma}}(t')\rangle\rangle$.
Other parts of $G_{\sigma\sigma}(t,t')$ vanish for $t'=t$, thus
are omitted as we are interested in $t'=t$ case. Furthermore, for
the sake of simplicity we will omit the real part in
$g_{k\alpha\sigma}^{r,a}$ which is justified in wide band limit.
Combining earlier obtained rate equations  with Eq.(\ref{eq12}) we arrive with the
rate equations expressed in Fourier space in the following form:
\begin{align}\label{eq15}
 \dot{\rho}_{00}=&-\frac{i}{2\pi}\int d\omega\sum_{\alpha\sigma}[\Gamma_{\sigma}^{\alpha}
 f^{\alpha}(\omega)G^>_{e\sigma\sigma}(\omega,\bar{t})
\nonumber \\
 &+\Gamma_{\sigma}^{\alpha}
 (1-f^{\alpha}(\omega))G^<_{e\sigma\sigma}(\omega,\bar{t})]
\nonumber \\
\dot{\rho}_{\sigma\sigma}=&\frac{i}{2\pi}\int
d\omega\sum_{\alpha}[\Gamma_{\sigma}^{\alpha}
 f^{\alpha}(\omega)G^>_{e\sigma\sigma}(\omega,\bar{t})
\nonumber \\
 &+\Gamma_{\sigma}^{\alpha}
 (1-f^{\alpha}(\omega))G^<_{e\sigma\sigma}(\omega,\bar{t})
 -\Gamma_{\bar{\sigma}}^{\alpha}
 f^{\alpha}(\omega)G^>_{d\sigma\sigma}(\omega,\bar{t})
 \nonumber \\
 &-\Gamma_{\bar{\sigma}}^{\alpha}
 (1-f^{\alpha}(\omega))G^<_{d\sigma\sigma}(\omega,\bar{t})]
\nonumber \\
\dot{\rho}_{22}=&\frac{i}{2\pi}\int
d\omega\sum_{\alpha\sigma}[\Gamma_{\sigma}^{\alpha}
f^{\alpha}(\omega)G^>_{d\bar{\sigma}\bar{\sigma}}(\omega,\bar{t})
\nonumber \\
&+\Gamma_{\sigma}^{\alpha}
(1-f^{\alpha}(\omega))G^<_{d\bar{\sigma}\bar{\sigma}}(\omega,\bar{t})]
\end{align}
%

%%Transition from time-space to Fourier space
As the transition from time space to the Fourier space in the time
dependent phenomena is not straightforward it is required to
justified it. Therefore, we introduced new time variables: a
mean time $\bar{t}=\frac{t+t'}{2}$ which varies slowly and a fast
varying time difference $\delta t=t-t'$, and expressed the Green
functions in these new time scales, i.e. $G(t,t')\longrightarrow
G(\delta t,\bar{t})$.\cite{davis,hernandez} Expanding $G(\delta
t,\bar{t})$ in the slow variable ($\bar{t}$) and taking the
Fourier transform with respect to the fast variable, we arrive at
the Green function $G(\omega ,\bar{t})=\sum_n\bar{G}^{(n)}(\omega
,\bar{t})\bar{t}^n$ with $\bar{G}^{(n)}$ being n-th derivative (of the $G(\omega ,\bar{t})$) with respect to the slow variable.
Then, we retain only the first term in this expansion
which allows us to write the lesser dot-leads Green function as in
Eq.(\ref{eq12}) This (lowest order)
gradient expansion is sufficient approach as we are
interested in sequential tunneling regime.\cite{davis}
After exploiting the above obtained equations, the rate equations acquire
form as these presented in Ref.[\onlinecite{dong}] when putting
intradot spin-flip parameter $R_{\sigma}$ to be equal to zero.
However, in the situation considered here, the dots' Green functions
depends on both $\omega$ and the mean time $\bar{t}$.
The dots' Green functions we find in the weak coupling
approximation, deriving them from corresponding equation of motion
for the dots operators. Technically, we assumed there is no coupling ($V_{i\sigma}^\alpha=0$)
and that leads are taken to be in local thermal equilibrium. Thus, we obtained:
\begin{align}\label{eq16}
G^<_{e\sigma\sigma}(\omega,\bar{t})&=i2\pi\rho_{\sigma\sigma}\delta(\omega-\epsilon_\sigma(\bar{t}))
\nonumber \\
G^>_{e\sigma\sigma}(\omega,\bar{t})&=-i2\pi\rho_{00}\delta(\omega-\epsilon_\sigma(\bar{t}))
\nonumber \\
G^<_{d\bar{\sigma}\bar{\sigma}}(\omega,\bar{t})&=i2\pi\rho_{22}\delta[\omega-(\epsilon_\sigma(\bar{t})+U)]
\nonumber \\
G^>_{d\bar{\sigma}\bar{\sigma}}(\omega,\bar{t})&=-i2\pi\rho_{\bar{\sigma}\bar{\sigma}}
\delta[\omega-(\epsilon_\sigma(\bar{t})+U)],
\end{align}
where the time dependence is clearly emphasized.
To derive these Green functions we used adiabatic approximation expanding $\epsilon_{\sigma}(t)$
around the mean time $\bar{t}$ and kept the terms up to linear
order in the slow variable, namely $\epsilon_{\sigma}(\tau)\approx
\epsilon_{\sigma}(\bar{t})+\dot{\epsilon}_{\sigma}(\tau)|_{\tau=\bar{t}}(\tau-\bar{t})$.
This allowed us to write
$\int_{t'}^{t}dt_1\epsilon_{\sigma}(\tau)\approx
\epsilon_{\sigma}(\bar{t})\delta t$. Then, after making Fourier
transformation, Eqs. (\ref{eq16}) are obtained.
Finally, connecting
Eqs.(\ref{eq16}) with Eqs.(\ref{eq15}) we arrive at the coupled set of differential equations
which we solve numerically to obtain time dependence of the density matrix elements.
%Again, the rate equations look like these in Ref.[\onlinecite{dong}].
%

Current flowing from $\alpha$ ($\alpha=S1,S2,D$) lead to the $j$th
dot is obtained from the standard definition:
\begin{eqnarray}\label{eq18}
J_\alpha^j=-e\langle\dot{N}_{\alpha}\rangle=-i\frac{e}{\hbar}\langle
[H,N_{\alpha}]\rangle,
\end{eqnarray}
where $N_\alpha$ is an occupation number operator in $\alpha$ lead.
After performing similar calculation as above, the current formula
becomes\cite{dong}:
\begin{align}\label{eq19}
J_\alpha^j&=i\frac{e}{\hbar}\int\frac{d\omega}{2\pi}\sum_{\sigma}\limits\Gamma_{j\sigma}^{\alpha}
f^{\alpha}(\omega)[G^>_{e\sigma\sigma}(\omega,\bar{t})+G^>_{d\bar{\sigma}\bar{\sigma}}(\omega,\bar{t})]
\nonumber \\
&+\Gamma_{j\sigma}^{\alpha}
[1-f^{\alpha}(\omega)][G^<_{e\sigma\sigma}(\omega,\bar{t})+G^<_{d\bar{\sigma}\bar{\sigma}}(\omega,\bar{t})].
\end{align}
Current passing through $j$th dot can be symmetrized in the
following way: $J^j=(J^{Sj}_j-J^D_j)/2$.
Total current flowing through the system is equal to $J=J^1+J^2$.
We assume that the distance between contacts in the drain lead is lesser than coherence length.
%%%%%%%%%%%%%%%%%%%%%%%%%%%%%%%%%%%%%%%%%%%%%%%%%%%
\section{Numerical Results}\label{Sec:3}
Time-dependent phenomenon is investigated in electronic transport
through two quantum dots coupled to external leads as shown
schematically in Fig.\ref{Fig:1}. The dots' levels are driven by
time-dependent gate voltages (ac force), whereas the dot-lead
couplings are assumed to be constant in time. Each dot has its own
gate electrode, thus the dots' levels can be tuned independently.
Moreover, this allows to apply AC voltages to two dots with
distinct external frequency and driving amplitudes. It is worth
nothing that this can not be achieved in multilevel single quantum
dot.

We consider the dots' levels driven by sinusoidal AC voltage, and
thus we assume $\epsilon_{i\sigma}(t)=\epsilon_{i\sigma}+\delta_i\cos{\Omega_i
t}$. Here, $\Omega_i$ is frequency, whereas $\delta_i$ is
amplitude of the external signal applied to the $i$th dot. In our
model, the chemical potentials of the source and drain leads are
set as $\mu_{S1}=\mu_{S2}=eV/2$ and $\mu_D=-eV/2$. Here, $V$ is a
bias voltage applied between the source (S1, S2) and drain leads.
Before the time dependent signals drive the dots' levels, the system
is in deep nonequilibrium due to applied bias voltage. Thus, one should
expect the dynamics of the system undergoes non-Marcovian processes.
In numerical calculations we assume that each dot is equally
coupled to its pair of leads, namely,
$\Gamma_{1}^{S}=\Gamma_{2}^{S}=\Gamma_{1}^{D}=\Gamma_{2}^{D}=\Gamma$
with $\Gamma$ being the energy unit.
Moreover, we assume spin degenerate and equal time-independent
parts of the dot levels, $\epsilon_{i\sigma}=\epsilon_0$ (for $i
=1,2$ and $\sigma =\uparrow ,\downarrow$) and equal amplitudes of
the oscillating signals ($\delta_1=\delta_2=\delta$). For
simplicity we also assume the same Coulomb parameters for the two
dots, $U_1=U_2=U$.

Approximations made during calculation of the rate equations and
current formula (gradient expansion and weak coupling) constrict
our model to special regimes. There are two regimes when this
approximation is valid: i) when $\epsilon_i\approx\mu_{Si}$ then must
be $\hbar\Omega\ll k_BT$, $\Gamma\ll k_BT$, ii) when
$-W/2<\epsilon_i<\mu_{Si}$ is valid for $\hbar\Omega\ll W$ and
$\Gamma\ll W$.\cite{davis} In our numerical calculations we choose
a set of parameters which fulfill these limitations. Moreover, we
assume that the system initially occupies the empty state
$\rho^{(1,2)}_{00}=1$. In our calculations we also set $h\equiv 1$.
\begin{figure}[t]
\begin{center}
  \includegraphics[width=0.9\columnwidth]{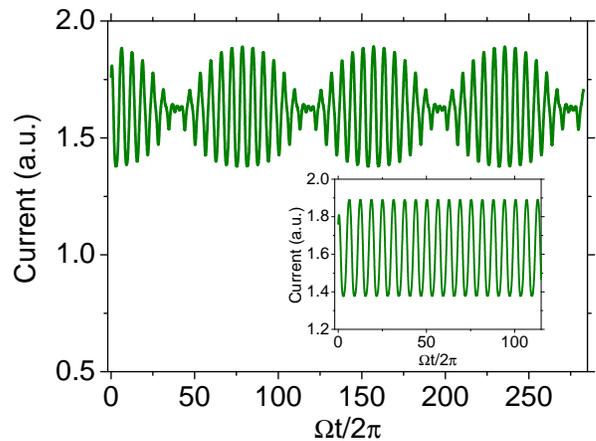}
  \caption{\label{Fig:2}
Current as a function of time calculated for nonmagnetic leads and
for different driving frequencies $\Omega_1=1.04\Omega$,
$\Omega_2=0.96\Omega$. The inset shows current evolution calculated for
nonmagnetic leads and for equal driving frequencies
$\Omega_1=\Omega_2=\Omega$.
Other parameters: $\epsilon_0=-\Gamma$, $\delta=0.2\Gamma$, $U=\Gamma$, $k_BT=0.1\Gamma$, $eV=2\Gamma$.
Here, $\Omega$ is chosen to be frequency unit. }
\end{center}
\end{figure}
\subsection{Nonmagnetic leads}\label{Sec:2A}
%Nonmagnetic leads
At the beginning we consider quantum dots coupled to nonmagnetic
leads and assume that dots' levels are driven by gate voltages
with different frequencies ($\Omega_1\neq\Omega_2$) but equal
amplitudes ($\delta_1=\delta_2$). When the external frequencies
differ only a little,  the beating in current are observed as shown
in Fig.\ref{Fig:2}. Total current beats with frequency being
twice the difference of the frequencies of the currents passing
through each quantum dot. Thus, the total current can be
decomposed as a product of two parts: one oscillates with the
average frequency $F=\frac{1}{2}(f_1+f_2)$ and second changing
with the frequency $\Delta f=\frac{1}{2}(f_1-f_2)$, where $f_1=\Omega_1/2\pi$
and $f_2=\Omega_2/2\pi$ are corresponding frequencies of the currents flowing
through each dot. The latter term controls the amplitude of the
\emph{envelope} and is responsible for the sensing of beating. The
beating frequency is twice the difference frequency $F_b=2\Delta f$.
Thus, the beating frequency is lowered when reducing the difference in the frequencies of the input
signals.
This effect is only due to the difference in the frequencies of
the external gate voltages. To show this we calculated the current
evolution for equal external frequencies $\Omega_1=\Omega_2$ and
displayed it in inset of Fig.\ref{Fig:2}. Thus, we believe that
this system is favorable for observing current's beating in
experiment. In contrast to results presented in
Ref.[\onlinecite{souza}], where the beating signal is damped (due to the
dot-lead coupling), in our case beating of the current is sustained in
time.

\begin{figure}[t]
\begin{center}
  \includegraphics[width=0.9\columnwidth]{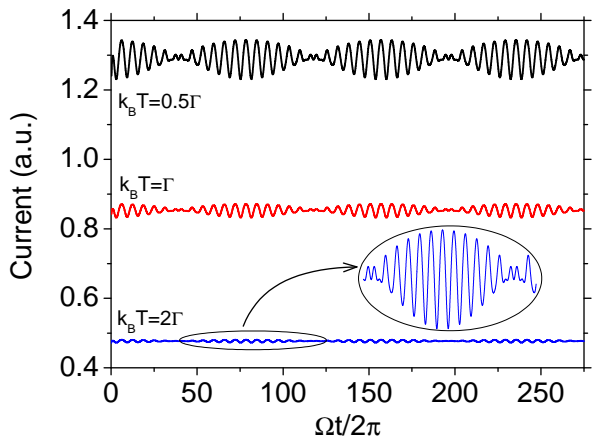}
  \caption{\label{Fig:3}
Beating current calculated for indicated values of the temperature.
Other parameters as in Fig.\ref{Fig:2}. In the ellipse we show zoomed part of the plot encircled in the
frame.}
\end{center}
\end{figure}

In Fig.\ref{Fig:3} we show the influence of the temperature on the current's beating. We notice that the amplitude of the beating signal is damped as temperature increases. However, even for $k_BT\gg\Gamma$ the beating pattern can still exist what is clearly shown in the zoomed part of Fig.\ref{Fig:3}. In turn, the amplitude of the beating can be increased by enlarging the amplitudes of the input signals ($\delta$). This implies that even for $k_BT>\Gamma$ the current's beating survives and may be observed when $\delta$ is sufficiently large. We also noticed that average current drops with increasing temperature, which is due to thermal damping effect in the leads.

\begin{figure}[t]
\begin{center}
  \includegraphics[width=0.9\columnwidth]{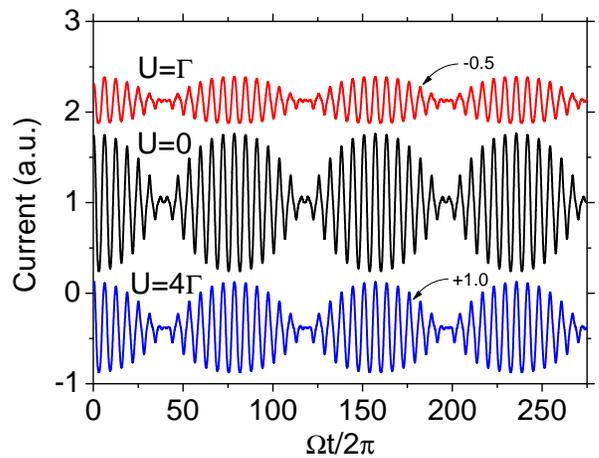}
  \caption{\label{Fig:4}
Beating current calculated for indicated values of the intradot Coulomb repulsion parameter.
Other parameters as in Fig.\ref{Fig:2}. Here, for clarity we rescaled the current plot for $U=\Gamma$ and $U=4\Gamma$. To obtain calculated values of the current for those parameter one has to add certain value to all points as is indicated by arrows.}
\end{center}
\end{figure}

Our calculations have also shown that intradot Coulomb interactions do not destroy beating pattern in current. To show this we plot in Fig.\ref{Fig:4} beating current for different values of the Hubbard parameter $U$. However, the Coulomb repulsion influences both the amplitude of the beating and the value of the average current. Namely, when there is no Coulomb interaction, the amplitude of the beating is most pronounced. When the on-dot Coulomb repulsion is present, the amplitude of the beating is suppressed. The dependence of the beating's amplitude is nonmonotonic function of the parameter $U$. When $\epsilon_0+U$ is within the transport window
it decays with increasing $U$, but for $\epsilon_0+U>\mu_S$ it starts to increase. However, it never again reaches the maximum value.

The average current is a also nonmonotonic function of the Coulomb parameter $U$.
It reaches high values when $\epsilon_0\approx\mu_D$ and $\epsilon_0+U$ approaches $\mu_{Sj}$
(but not very close, $\epsilon_0+U\ncong\mu_{Sj}$, due to imposed gradient's expansion condition.) When $\epsilon_0+U$ is
beyond the transport window, average current drops and saturates for sufficiently large $U$.
Moreover, Coulomb interactions introduce small horizontal asymmetry in the beating pattern.
%Ferromagnetic leads
%
\begin{figure}[t]
\begin{center}
  \includegraphics[width=0.56\columnwidth]{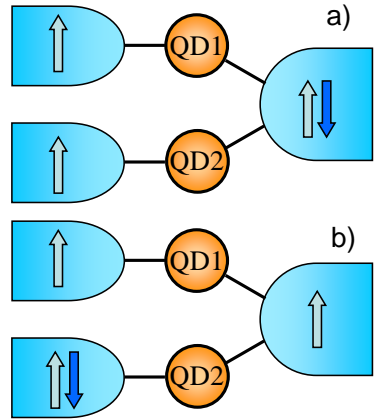}
  \caption{\label{Fig:5}
Magnetic configurations taken into account.}
\end{center}
\end{figure}
\subsection{Ferromagnetic leads}
When, the leads are ferromagnetic several magnetic configurations
are possible. To 'measure' the difference in these distinct
configurations it is convenient to introduce tunnel
magnetoresistance (TMR). This quantity results from spin-dependent
dot-lead tunneling processes, which in turn leads to the
dependence of transport characteristics on magnetic configuration
of the system. The TMR is quantitatively described by the ratio
${\rm TMR}=I_P-I_{AP} / I_{AP}$, where $I_P$ and $I_{AP}$ denote the
currents flowing through the system in the parallel and
antiparallel magnetic configurations, respectively.
\begin{figure}[t]
\begin{center}
  \includegraphics[width=0.9\columnwidth]{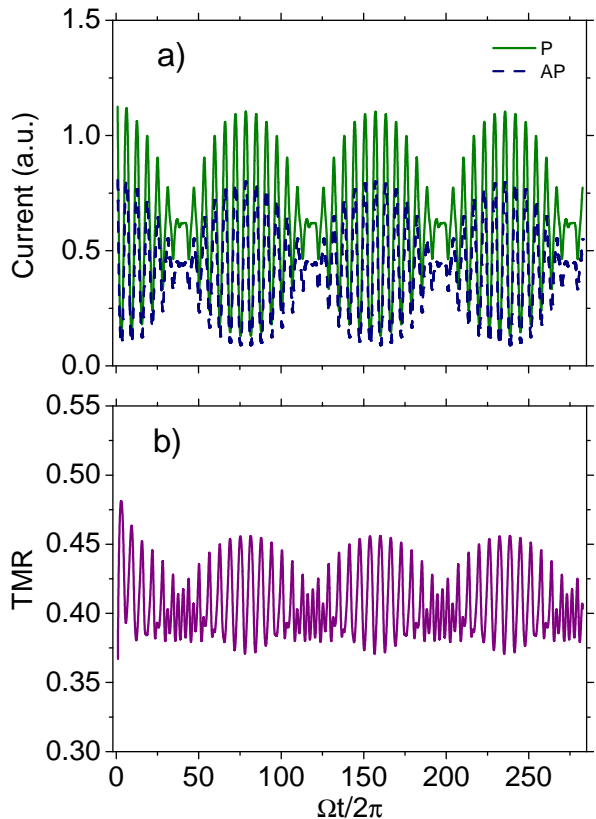}
  \caption{\label{Fig:6}
a) Current as a function of time calculated for two magnetic
configurations from Fig.\ref{Fig:5}(a) and for $p=0.5$, $U=4\Gamma$. b)
Time evolution of TMR. Other parameters as in Fig.\ref{Fig:2}.}
\end{center}
\end{figure}
Introducing the spin polarization $p_{\alpha}$ of lead $\alpha$
($\alpha=S1,S2,D$) as $p_{\alpha}=(\rho_{\alpha}^{+}-
\rho_{\alpha}^{-})/ (\rho_{\alpha}^{+}+ \rho_{\alpha}^{-})$, the
coupling parameters can be expressed as
$\Gamma_{i\alpha}^{+(-)}=\Gamma_{i\alpha}(1\pm p_{\alpha})$, with
$\Gamma_{i\alpha}= (\Gamma_{i\alpha}^{+}
+\Gamma_{i\alpha}^{-})/2$. Here, $\rho_{\alpha}^{+}$ and
$\rho_{\alpha}^{-}$ are the densities of states at the Fermi level
for spin-majority and spin-minority electrons in the lead
$\alpha$, while $\Gamma_{i\alpha}^{+}$ and $\Gamma_{i\alpha}^{-}$
describe coupling of the $i$th dot to the lead $\alpha$ in the
spin-majority and spin-minority channels, respectively.

\begin{figure}[t]
\begin{center}
  \includegraphics[width=0.9\columnwidth]{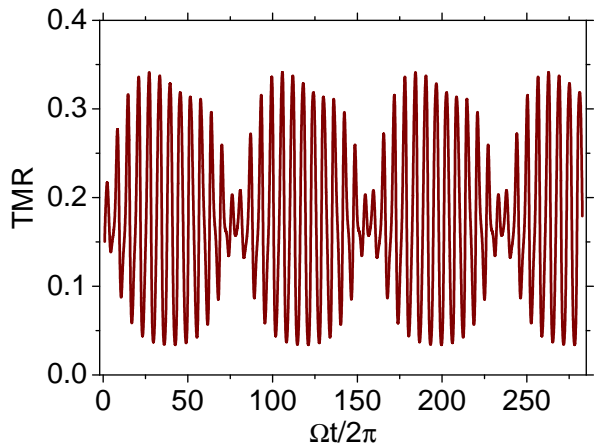}
\caption{\label{Fig:7} Time dependence of the TMR calculated for
magnetic configurations from Fig.\ref{Fig:5}(b). Other parameters
as in Fig.\ref{Fig:6}.}
\end{center}
\end{figure}
Let us first consider the case where the magnetic moments of the
source leads are pinned (with 'up' direction) and the
magnetization of the drain electrode can be changed from 'up' to
'down' as schematically is shown in Fig.\ref{Fig:5}(a). We
calculated the currents in both magnetic configurations and TMR
for leads' polarization $p_{S1}=p_{S2}=p_D=0.5$. Firstly, one
observes that the beating is still present in the current
characteristics for both magnetic configurations. However, TMR
exhibits beating pattern a little distorted.
As the dots are decoupled from each other and we consider the case of
weak couplings we should expect positive TMR, which is clearly
displayed in Fig.\ref{Fig:6}(b). Then, off course, the current
in parallel magnetic configuration is greater than that in
antiparallel one (see Fig.\ref{Fig:6}(a)). However, spin symmetry breaking
processes, as spin-flip scattering, may change the sign of TMR as shown
in Ref.[\onlinecite{stefanucci2}]. Here, we do not consider such processes.
Recent experiments have shown that the spin relaxation time in quantum dots
can reach millisecond\cite{elzerman,johnson,koppens} or even second timescales\cite{amasha}
which is much longer than electron tunnel rate ($\sim\Gamma^{-1}$).

Now, we consider the
situation when the magnetization of the drain lead and one of the
source electrode are pinned, whereas the magnetic moment of the
second source lead can be flipped, as is shown in Fig.\ref{Fig:5}(b).
In this case, the difference between 'parallel' and 'antiparallel'
configurations is less visible, which results in suppression of
TMR. However, the oscillating character is still conserved and
here the beating is even more pronounced (see Fig.\ref{Fig:7}).
The suppression of the TMR for this magnetic configurations is clear when one notices
that for this case only one transport channel is partially blocked
(due to relevant difference in the orientations of the leads
magnetic moments in the AP configuration), whereas in the former
case both channels are bad 'conductors' in the AP configuration.
Moreover, in this magnetic configuration, $\pi/2$ phase shift is induced
in the TMR pattern.
A small distortion in the upper semicircles comes from the different symmetries
of the current profiles for P and AP configurations in the vicinity of the \emph{node} points.

\subsection{Spin-biased leads}
Here, we consider the double  dots' system subjected to the source and drain spin
batteries\cite{hirsch,brouver,wang2} which provides pure spin current without accompanying charge current.
Pure spin current is one of the most important points for spintronics. However, so far spin control methods
in commercial devices mainly have relied on usage of magnetic field\cite{mucciolo,watson} or optical techniques which are not very
efficient. Recent experiments show that pure spin current can be all-electrically generated in a
micron-wide channels of a GaAs two-dimensional electron gas\cite{frolov1,frolov2}. This is very important from
the application point of view, because other quantum systems can be
easily integrated with such all-electrically controllable
spin battery. To control such a device we do not need optical or
magnetic fields which precisely adjusting is rather great effort and thus, useless for commercial applications.
\begin{figure}[t]
\begin{center}
  \includegraphics[width=0.9\columnwidth]{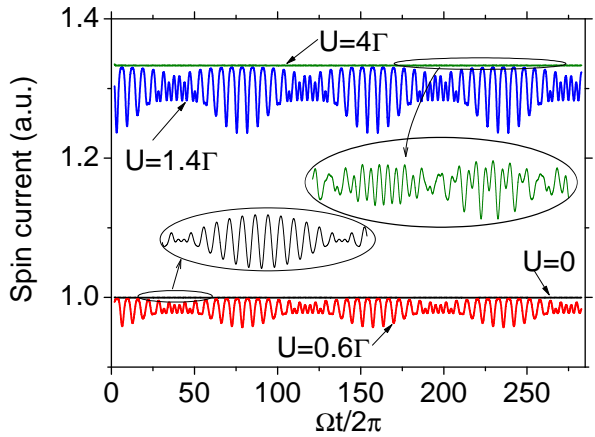}
\caption{\label{Fig:8} Time evolution of the spin current calculated for indicated values of the
intradot Coulomb repulsion parameter and for $\epsilon_0=0 $ in case of the symmetric spin batteries. Other parameters
as in Fig.\ref{Fig:2}.}
\end{center}
\end{figure}

First, we investigate DQD system connected to symmetric dipolar spin batteries, i.e. we assume that
$\mu_{Sj\uparrow}=\mu_{D\downarrow}$ and $\mu_{Sj\downarrow}=\mu_{D\uparrow}$ for ($j=1,2$).
Introducing the spin bias $V_s$, generally we may write $\mu_{Sj\sigma}=e(V+\tilde{\sigma}V_s)/2$ and $\mu_{D\sigma}=-e(V+\tilde{\sigma}V_s)/2$
with $\tilde{\sigma}=1$ ($\tilde{\sigma}=-1$) for $\sigma=\uparrow$ ($\sigma=\downarrow$)\cite{swirkowicz}.
As we are interested in pure spin current we further set bias voltage equal to
zero $V=0$. In this case the net charge current vanishes, because all spin-up electrons
flow in one direction and equal amount of spin-down electrons flow in the opposite direction, and only
pure spin current is generated. The spin current is defined in the following way, $J_s=(J_{\uparrow}-J_{\downarrow})/e$, where
$J_{\sigma}=J_{\sigma}^1+J_{\sigma}^2$ ($\sigma=\uparrow,\downarrow$). However, it is worth to mention that
when the dot's energy level is split
i.e., $\epsilon_{\uparrow}\neq\epsilon_{\downarrow}$, the nonzero charge current can be generated\cite{wang2,bao,swirkowicz}.
In Fig.\ref{Fig:8} we show
time evolution of the spin current calculated for different strengths of the intradot Coulomb interactions.
One can notice that the spin current exhibits more complicated beating pattern (similar as TMR in Fig.\ref{Fig:6}).
On the other hand, for noninteracting case ($U=0$) we notice the clear evidence of pure beats in the spin current.
However, for this case the amplitude of the beating is small. When Coulomb interactions are turned on, symmetric beating
pattern vanishes and even more features appear. As intradot repulsion increases, the beating in the spin current
become more and more asymmetric and the \emph{node} points cease to exist. Instead of \emph{node} new oscillations emerge.
Namely, for nonzero $U$, spin current evolution composes from two kind of oscillations: main oscillations and some
sub-oscillations emerged in the vicinity of the \emph{nodes} points (existing in noninteracting case).
For sufficiently large $U$,
the main oscillations become very asymmetric and the sub-oscillations are more pronounced.
In contrast to the nonmagnetic case, the dependence on the Coulomb repulsion is here much more complex.
The amplitude of the beating is small for both small and large enough $U$. This is because the state $\epsilon_0+U$ is
far away from the chemical potentials of the leads. However, when $\epsilon_0+U$ is outside the transport window,
the average spin current grows meaningly (in contrast to the charge current in nonmagnetic case from Fig.\ref{Fig:4}).
When $U$ is sufficiently large (i.e. $\epsilon_0+U\gg\mu_{\alpha\sigma}$) the probability of double occupancy
drops almost to zero ($\rho_{22}\approx 0$) and the occupation numbers $n_\sigma$ also decrease
(however $\rho_{\sigma}$ increases). This enables effectively faster tunelling processes through QD and thus
enlarges the spin current. For $U\gg\mu_{\alpha\sigma}$ the spin current becomes saturated.
\begin{figure}[t]
\begin{center}
  \includegraphics[width=0.9\columnwidth]{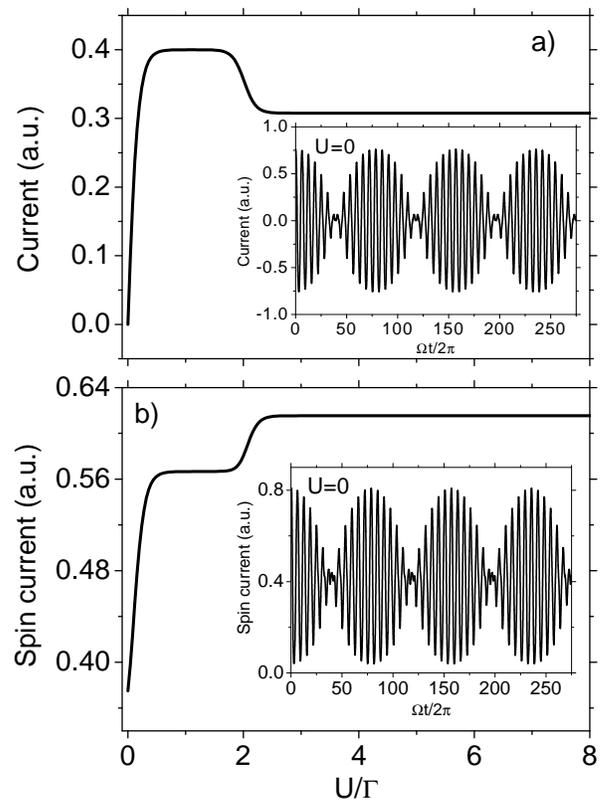}
\caption{\label{Fig:9} Stationary charge (a) and spin (b) current as a function of the
Coulomb repulsion parameter $U$.
Insets: Time evolution of the charge and spin current,respectively, calculated for for $\epsilon_0=0$ and for $U=0$,
in case of the asymmetric spin batteries. Other parameters
as in Fig.\ref{Fig:2}.}
\end{center}
\end{figure}

Now, we consider DQD system attached to asymmetric spin batteries. In this case we set
$\mu_{Sj\sigma}=\tilde{\sigma}eV_s$ and $\mu_{D\sigma}=0$. Let us first consider noninteracting case
($U=0$). The dots' energy levels are
situated symmetrically with respect to the spin bias voltages of the source leads, e.g.,
$\epsilon_0$ is in the mid between $\mu_{Sj\uparrow}$ and $\mu_{Sj\downarrow}$. Thus, the same amount
of spin-up electrons
flows in one direction and equal amount of spin-down electrons flows in the opposite direction and the average spin
current is nonzero. However, due to oscillations of the dots' levels, the charge current is also generated, but on average
it vanishes. In the case of asymmetric spin batteries both the spin and charge current exhibits well-defined beating
pattern as shown in the insets of Fig.\ref{Fig:9}.
When the Coulomb interactions are turned on, a nonzero average charge current is induced.
This is because earlier mentioned symmetry is now broken.
It is worth noting that such symmetry exists also when $\epsilon_0=-U/2$.
However, for $\epsilon_0<\mu_D$ and sufficiently
large $U$ the system is in the Coulomb blockade, thus, we expect zero current
in the weak coupling regime.
Firstly, for a small value of the parameter $U$, the average charge current grows very fast reaching
maximum value for $U\approx 0.4\Gamma$ and then is unchanged with further increase
of the $U$, until $\epsilon_0+U$ exceeds $\mu_{Sj\uparrow}$, when it becomes reduced a little and
saturates.
This drop in average charge current is because the state $\epsilon_0+U$ ceases to contribute to the transport.
In turn, the average spin current, generally, grows with increasing parameter $U$ (regardless a certain ranges of $U$ where the
average spin current is constant). When $U$ is sufficiently large the average
spin current is also saturated. To show this dependencies we plotted stationary charge and spin current
in Fig.\ref{Fig:9} which may be regarded as average values of respective currents in time-dependent phenomenon.
However, one should bear in mind that this is not true for $\epsilon_0+U$ being close to $\mu_{S\uparrow}$ due to
gradient expansion condition. Hence, this range should not be disregarded.

It is also worth noting that for $\epsilon_0<\mu_D$ both the average charge and spin current can
change the sign. As a result one should expect negative charge and spin
differential conductances. Moreover, in contrast to the symmetric spin batteries, here, the beating structure in spin current is very symmetric for all values
of the Coulomb interactions parameter $U$.

\section{Final conclusions}\label{Sec:4}
%some remarks....
In summary, we have studied coherent transport through two
uncoupled quantum dots, which are attached to nonmagnetic and/or
ferromagnetic leads. Generally, two magnetic configurations were
discussed. We took into account the Coulomb interaction between
electrons on the same dot and calculated transport characteristics
in the nonlinear response regime, using the rate equation approach
connected with Green functions method and with slave-boson
formalism. Our analysis was performed for oscillating dots' energy
levels within the gradient expansion approximation.

We have found clear evidence of both charge and spin current beating as well as the
beating pattern in TMR. We have shown that the effect is due to
the difference in the frequencies of the applied gate voltages to
the two dots. In magnetic case, beating in spin current or TMR
may be deformed. However, for DQD system coupled to the
asymmetric spin battery spin current exhibits well-defined beating
structure.

In this paper we have omitted the interdot Coulomb repulsion as in real systems it is
much more smaller than the intradot Coulomb interactions.
Moreover, for the parameters assumed in this paper the interdot Coulomb
integral\cite{dong2} would be (much) lesser than the dot-lead coupling strength,
and that's why it does not lead to the splitting in the dot's density of state.
Correspondingly, sufficiently small interdot interaction does not affect considered phenomenon and
is irrelevant. However, sufficiently strong interdot Coulomb interaction can introduce some deviation in the beating
pattern.

%some applications
The proposed system can be used as a device to measure frequency
of an unknown signal. Then, one needs only one QD in one arm
coupled to the source and drain leads, whereas the second arm
delivers the unknown signal. The arm with QD plays role as the
reference channel, and thus, tuning the frequency of the reference
signal one is enable to detect the frequency of the 'unknown'
signal.
%%%%%%%%%%%%
Moreover, the DQD device presented above can be utilized in coding
information (signal). Thus, such device may be called nanoscale
superheterodyne. Using such a device we are able to mix two signals
of slightly different frequencies. As a result,
one obtain resultant signal being a composition
of slow-varying and fast-varying parts (as mentioned in Sec.\ref{Sec:2A}).
Then, one of the signals with, for instance low frequency, may be extracted and further
processed. The advantage of the device is that a signal with lower frequency is easier to be processed.

\begin{acknowledgements}
The author thanks J. Barna\'s
and K. Walczak for helpful discussions.
This work, was supported partly by funds from the Polish Ministry
of Science and Higher Education as a research Project No. N202
169536 in years 2009-2011. The author also acknowledges support by
funds from the Adam Mickiewicz University Foundation.
\end{acknowledgements}

\end{document}